**A high-resolution carbon balance in a small temperate catchment: insights from the Schwabach River, Germany**

Kern Y. Lee*, Robert van Geldern, Johannes A.C. Barth

Friedrich-Alexander University Erlangen-Nuremberg (FAU), Department of Geography and Geosciences, GeoZentrum Nordbayern, Schlossgarten 5, 91054 Erlangen, Germany.

* Corresponding author. E-mail: kern.lee@fau.de

**Abstract**

The present study is one of only few to address the stable carbon isotope ($\delta^{13}$C) and concentration dynamics of DIC, DOC, and POC over an entire hydrologic year, using a dataset of high sampling resolution (4 to 11 samples retrieved per month). This research was performed in the catchment of the Schwabach River, a typical mid-latitude small headwater stream in Germany emplaced mainly in karstic bedrock. The DIC data indicated the dominance of mineral weathering as a DIC source, with a noticeable dilution effect during periods of high river flow. A weakly negative relationship between discharge and $\delta^{13}C_{DIC}$ hints at a greater importance of plant-derived organic matter during flooding events, likely transported to river waters via overland runoff and intermediate flow. DOC inputs appeared to be enhanced during periods of high discharge, indicating a greater importance of overland runoff as a DOC source. POC concentrations seem unaffected by changes in discharge, although a slight negative correlation between $\delta^{13}C_{POC}$ and discharge may be derived from increased inputs of C4 plant material.

Estimated $CO_2$ concentrations were in excess of ambient atmospheric values throughout the year, confirming that the surface waters of the Schwabach River are a net $CO_2$ source. The total riverine carbon flux was dominated by DIC (70%), followed by $CO_2$ outgassing (21%), DOC (7%), and POC (2%). While the selection of a bi-monthly sampling scheme yielded a broadly similar carbon flux estimate to that utilizing the entire dataset, the use of a monthly sampling interval differed by as much as 19% from values using the high-resolution data set. This discrepancy is due to the inability of a monthly sampling scheme to capture sudden and large variations in river discharge and associated changes in dissolved/particulate carbon concentrations, such as those observed during flooding. We suggest that bi-monthly sampling is the minimum timeframe required to achieve an acceptable degree of accuracy in the calculation of carbon fluxes, at least during periods of high runoff. The application of high sampling frequencies and comprehensive DIC, DOC, and POC studies in future research would greatly reduce uncertainties in local riverine carbon budgets, and help clarify the role of smaller streams in the global carbon cycle.

## 1. Introduction

Riverine fluxes of terrestrial carbon to the ocean represent a major link in the global carbon cycle, and much research has focused on clarifying fluvial carbon budgets worldwide (Aufdenkampe et al., 2011; Battin et al., 2009; Cole et al., 2007; Hope et al., 1994; Sawakuchi et al., 2017). However, watershed carbon flux estimates are subject to large uncertainties, of which an incomplete knowledge of riverine carbon dynamics plays a large part (Meybeck, 1993; Raymond et al., 2013; Wehrli, 2013). While the primary importance of rivers in carbon cycling lies in their linkage of terrestrial and oceanic reservoirs, carbon within river waters are also subject to in-situ production, consumption, and loss (Cole et al., 2007; Orozco-Durán et al., 2015; Ward et al., 2017). Complex interrelationships between these processes, and carbon phase changes within river waters, can make riverine carbon budgets difficult to constrain.

This gap in knowledge is even greater with regard to smaller rivers, particularly headwater streams, for which information on regional fluvial carbon cycles is sparse. However, headwater streams play an important role in regional and global carbon cycle dynamics as they are the primary source of groundwater and soil-derived carbon. Several studies suggest that particularly large amounts of $CO_2$ evade to the atmosphere by networks of smaller streams and rivers due to elevated DIC concentrations, higher gas transfer velocities, and the large cumulative global surface areas encompassed by lower-order streams (Lauerwald et al., 2015; Lee et al., 2013; Raymond et al., 2013). However, despite the relative significance of $CO_2$ evasion in small rivers and streams, the quantification of these contributions remains largely uncertain (Cole et al., 2007).

River waters contain both organic and inorganic carbon, either in dissolved or particulate form. Dissolved inorganic carbon (DIC) can be sourced from carbonate or silicate weathering, *in-situ* biological respiration, or soil-respired carbon. In addition to DIC transport and sediment storage, a large proportion of aquatic DIC is lost as $CO_2$ before river waters reach the oceans. This carbon loss has been estimated to account for as much as 74% of total riverine carbon export (Raymond et al., 2013; Wehrli, 2013). Such $CO_2$ losses by rivers are an important link towards a better understanding of catchment productivity (Aufdenkampe et al., 2011; Butman and Raymond, 2011; Raymond et al., 2013).

Dissolved organic carbon (DOC) is usually present in the form of organic acids or humic material, being predominantly composed of soil leachate (Meybeck, 1993). In addition, some studies indicate that DOC can be exudated directly from living aquatic plants (Demarty and Prairie, 2009). While terrestrially-derived DOC may initially be recalcitrant in soils, it is more labile in well-lit and oxygenated river waters mostly due to in-situ biological processing and photo-oxidation (Dusek et al., 2017; Hedges et al., 1997), and reduced amounts of DOC are ultimately transported to the oceans. This DOC processing creates $CO_2$ as a by-product, contributing to the riverine DIC pool and further enhancing $CO_2$ outgassing (Mayorga et al., 2005)

The clarification and quantification of riverine carbon fluxes is an important facet of global carbon cycling research, but the accuracy of flux determinations primarily depends on sampling frequency. This is particularly true in regions where seasonal variations in climate are more variable, such as in temperate environments. Temporal and financial constraints can restrict the determination of sampling schedules, and many studies are limited to monthly sampling regimes due to these limitations. While this may be adequate for nominal flow and average climatic backgrounds, the occurrence of variable runoff conditions, including floods and droughts, could introduce hitherto unknown errors into annual carbon flux estimations. In particular, flooding events are not only responsible for large increases in lateral transport, but also changes in the predominant DOC and DIC inputs due to enhanced overland runoff (Bouillon et al., 2012; Cai et al., 2015; Geeraert et al., 2017)

When considering entire catchment carbon balances, it is also important to identify sources and sinks of carbon in addition to constraining flux estimates. This remains a challenge due to the complexities of riverine carbon cycling and interrelationships between all carbon phases. In addition to concentration measurements of the various carbon phases, stable carbon isotope ratios ($^{13}C/^{12}C$) can be useful tools in this regard, as production, processing, and fate in biogeochemical cycles often produce typical isotopic changes. These can vary depending on the type of carbon phases and processes involved. For example, DIC derived from marine carbonates usually have $\delta^{13}C$ values that average around 0 ‰, while atmospheric $CO_2$ currently has a value of about -8 ‰ (Clark and Fritz, 1997; Ghosh and Brand, 2003).

On the other hand, plant tissues exhibit a $^{13}C$-depletion of varying degrees relative to atmospheric $CO_2$. Depending on the photosynthetic pathway, this results in plant $\delta^{13}C$ values that range from averages of -12 ±1.1‰ (C4) to -26.7 ± 2.3‰ (C3) (Cerling et al., 1997; Ehleringer and Cerling, 2002). These isotopic differences are reflected in the $\delta^{13}C$ of particulate organic carbon (POC) and DOC derived from plants. When this organic carbon is turned over through organic matter decomposition and oxidation, the resulting DIC becomes depleted in $^{13}C$.

Note that DOC can also undergo $^{13}C$-enrichment due to processes such as photo-oxidation, and biological uptake and decomposition (Cai et al., 2015; Rounick and Winterbourn, 1986; Stanley et al., 2012). This results in a different input signal to the DIC when organic matter is turned over. Moreover, the $\delta^{13}C$ of DIC can also be subject to modification by $CO_2$ outgassing or aquatic plant uptake. Both processes enrich the residual DIC in $^{13}C$ (Barth et al., 1998; Doctor et al., 2008; Michaelis et al., 1985; O'Leary, 1988).

With these processes and complexities in mind, the pairing of well-constrained carbon flux estimates with $\delta^{13}C$ data can be a powerful tool in the elucidation of riverine carbon fluxes. This study uses a high temporal-resolution dataset of carbon river data spread out over one hydrological year to calculate DIC, DOC, POC, and $CO_2$ fluxes from the Schwabach, a small river in Northern Bavaria, Germany, with a catchment area of less than 200 $km^2$. The study is also useful because thus far, few studies have simultaneously analyzed concentrations and $\delta^{13}C$ of DIC, DOC, and POC (Cai et al., 2015; Cartwright, 2010; Geeraert et al., 2017; Hossler and Bauer, 2013). In addition, hardly any studies of this nature exist with a daily to weekly sampling frequency over the course of at least one hydrological year.

In this work, we calculate daily to weekly carbon fluxes in order to estimate total carbon export from a headwater stream in a mixed karstic and sandstone watershed. We also aim to determine the influences of sampling frequency on the accuracy of carbon flux estimates. For comparison and to determine the source or sink nature of the catchment, this carbon export is then compared to the net carbon input within the studied watershed, as calculated using satellite-derived NPP datasets. A further objective is to use the carbon concentration and stable isotope data to constrain carbon sources and sinks of the phases DIC, $CO_2$, DOC, and POC, and to investigate if seasonal changes influence the dominance of a particular carbon phase.

## 2. Materials and Methods

### 2.1. Site Description and Sampling

The Schwabach River Catchment is situated in Northern Bavaria, Germany, within the region of the Fraconian Alp. This basin encompasses 191 km$^2$, with the Schwabach River being 32.4 km in length (Bavarian State Office of the Environment). Waters were sampled at the lower river course from a site located in the city of Erlangen near the river's mouth, between June of 2013 and May of 2014 (Fig. 1). Sampling intervals were 4-5 samples taken during all months on at least a weekly basis. In addition, 11 samples were retrieved during June 2016 with up to a daily interval. Samples were collected more frequently during June of 2016 in order to capture changes during a period of exceptionally high water level, in contrast to the closer to nominal flow conditions that were observed during all other sampling periods.

Water from the Schwabach River was retrieved from a bridge with a bucket, which was rinsed out three times with river water prior to sample storage. The parameters of pH, water temperature (*T*), electric conductivity (EC) and dissolved oxygen (DO) were measured on-site either with a WTW Multi 350i (WTW GmbH, Weilheim, Germany) or a HQ 40d (HACH Company, Loveland, CO, USA) handheld multi parameter instrument, while total alkalinity (TA) was determined with a portable HACH Digital Titrator (Model 16900, HACH Company, Loveland, CO, USA).

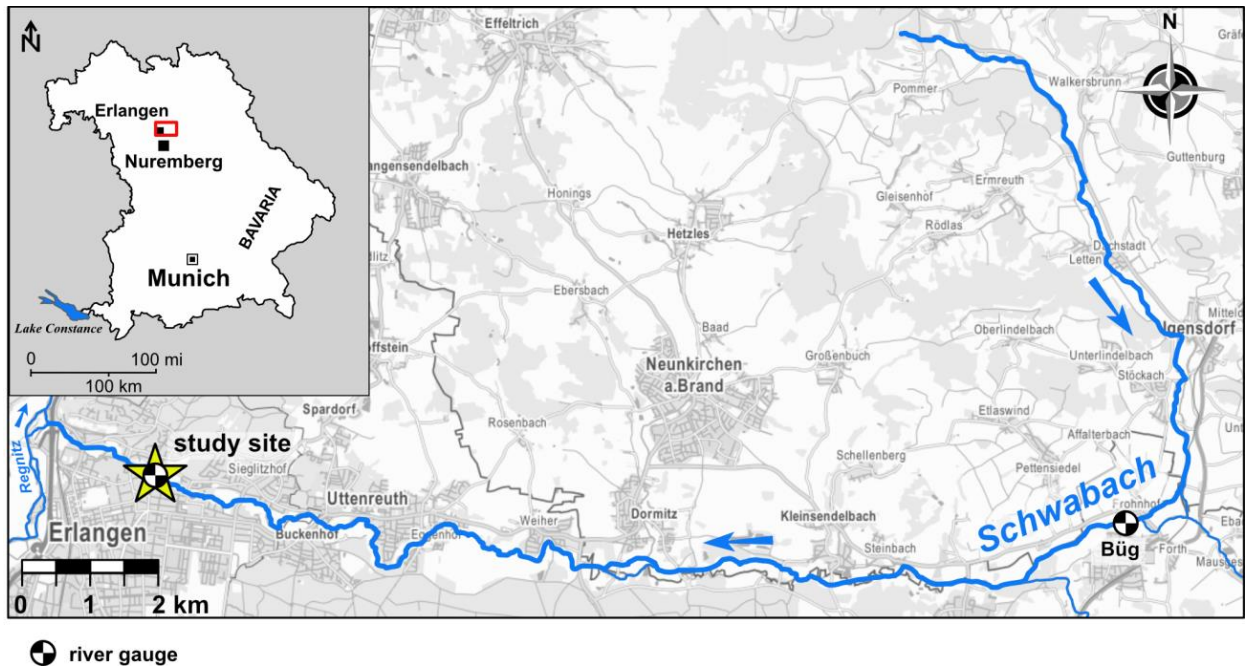

Figure 1. Map of the Schwabach River, showing the sampling site (yellow star), river gauging stations (chequered circles), and location of the watershed within Bavaria (map insert) Maps are modified from the following sources: Bavaria–Wikimedia Commons; Schwabach River–Bayernatlas (https://geoportal.bayern.de/bayernatlas/ (26.06.2017))

Samples for DIC and DOC concentration and isotope analyses were collected from 0.45 μm pore size filters (Minisart HighFlow PES, Sartorius AG, Germany) into 40 mL amber glass vials that met the standards of the US Environmental Protection Agency (so-called EPA vials). The vials were filled gently to the brim with no headspace to avoid air bubbles, and subsequently capped by pierceable caps with butyl rubber/PTFE septa. To ensure gas tightness, the septa were flipped, i.e. the butyl rubber side was positioned facing the sample. Beforehand, the EPA vials were poisoned with a few drops of nearly saturated $HgCl_2$ solution (~50 g $L^{-1}$) to prevent microbial alteration of the carbon species during sample storage. Samples were then stored in a cool and dark environment (4 °C) until analysis.

*2.2. Laboratory Analyses*

Water samples were analyzed for DIC and DOC concentrations, and their respective carbon stable isotope ratios $\delta^{13}C_{DIC}$ and $\delta^{13}C_{DOC}$, by an OI Analytical Aurora 1030W TIC-TOC analyzer (OI Analytical, College Station, Texas, USA) coupled in continuous flow mode to a Thermo Scientific Delta V plus (ThermoFisher Scientific, Bremen, Germany) isotope ratio mass spectrometer (IRMS). The sample was reacted with 1 mL of 5% phosphoric acid ($H_3PO_4$) at 70°C for 2 min to release the DIC as $CO_2$. The evolved $CO_2$ was purged from the sample by helium. In a second step, 2 mL of 10% sodium persulfate ($Na_2S_2O_8$) were reacted for 5 min at 98°C to oxidize the DOC to $CO_2$ and purged from the solution by helium. A trap and purge (T&P) system was installed for the analysis of low concentrations.

The data were normalized to the VPDB scale by two laboratory reference materials ($C_4$ sugar and KHP) measured in each run. Details of the coupling of the TIC/TOC analyzer to IRMS are described in St‐Jean (2003) and van Geldern et al. (2013); (Hélie et al., 2002). Concentration was determined from the signal of the OI Aurora 1030W internal non-dispersive infrared sensor (NDIR) and a set of calibration standards with known concentrations prepared from analytical (A.C.S.) grade potassium hydrogen phthalate (KHP). Areas of the sample peaks are directly proportional to the amount of $CO_2$ generated by the reaction of the sample with acid (DIC) or sodium persulfate (DOC). The precision based on analyses of in-house reference materials, expressed as standard deviation ($\pm 1\sigma$), was better than ±0.3‰ for $\delta^{13}C$. For concentration analyses the precision, defined as the standard deviation ($\pm 1\sigma$) of a control standard during the runs, was better than 5% relative standard deviation (RSD).

POC was collected onto pre-heated (400°C for 4 hours) glass fibre filters (GFF) with a pore size of 0.7 μm, which were subsequently dried for 8 hours. The entire GFF with captured and dried material was then pulverized by a ball mill (Retsch CryoMill) and the resulting fine powder fumigated by concentrated HCl in a desiccator for 24 hours to eliminate possible carbonate particles on the filter paper. Afterwards, the sample was stored for 2h at 70°C to degas any remaining acid fume. Samples were analyzed for $\delta^{13}C$ using a Costech Elemental Analyzer (ECS 4010; Costech International, Pioltello, Italy; now NC Technologies, Bussero, Italy) in continuous flow mode coupled to a Thermo Scientific Delta V plus IRMS. Analytical precision was better than 0.1‰ for $\delta^{13}C$. For carbon concentration analysis (% carbon), the relative standard deviation (RSD) was better than 2% (±1σ) based on the repeated analyses of a control sample (acetanelide).

*2.3. pCO₂ and Carbon Flux Calculations*

The DIC, pH and water temperature data were used to calculate partial pressures of $CO_2$ ($pCO_2$), according to the following relationship (Marx et al., 2017):

$$pCO_2 = \frac{HCO_3^- \times H^+}{K_H \times K_1}$$

(1)

where $K_1$ and $K_H$ are the temperature-dependent dissociation constants for $HCO_3^-$ and Henry's gas constant in units of mol L$^{-1}$ atm$^{-1}$, respectively. The dissociation constants were obtained from equations regressed by Clark and Fritz (1997) and Drever (1997).

These $pCO_2$ estimates were incorporated into another equation (Liss, 1973) to obtain efflux rates with:

$$F = k\left[(C_{water} - C_{atm}) \times \frac{1}{RT}\right]. \tag{2}$$

In equation (2), $F$ is gas efflux in units of g C m$^2$ day$^{-1}$, $k$ is the gas transfer velocity in m day$^{-1}$, C$_{water}$ and C$_{atm}$ are $CO_2$ partial pressures in the water and overlying air (assuming a global value of 397 ppmv, based on data averaged over the time period of the study that were retrieved from the National Oceanic and Atmospheric Administration (NOAA))[1], $R$ is the ideal gas constant with 8.2057 x$10^{-5}$ $m^3$ atm $K^{-1}$ $mol^{-1}$, and $T$ is temperature in ° Kelvin. The coefficient $k$ was calculated as $k_{600}$ (i.e., $k$ at a Schmidt number of 600) estimated from stream velocity (V) and slope (S) after Raymond et al. (2012):

$$k_{600} = V \times S \times 2841 + 2.02 \tag{3}$$

The parameters V and S were calculated from river flow data that were available in hourly to daily intervals, obtained from the Bavarian Environmental Agency[2] for the gauging stations Erlangen (station ID 24238501) and Büg (station ID 24238002). The latter is located 18 km upstream of the study site (Fig. 1). The areal $CO_2$ efflux estimates were then multiplied by the estimated water surface area of the Schwabach River and its major tributaries. These estimates were performed from Google Maps satellite imagery and river data from the Bavarian Environmental Agency[3].

River flow and geochemical data were also used to calculate yearly DIC, DOC, and POC fluxes. This was achieved by multiplying daily flow rates averaged over the time period to the next sampling event by measured daily to weekly concentrations to establish average monthly values. These values were used in turn to obtain values averaged over an entire year (in units of kt C $yr^{-1}$). Flux estimates based on the complete dataset were compared to those calculated on the assumption of bi-monthly and monthly sampling, in order to determine the extent to which different sampling resolutions may affect the accuracy of carbon export determinations.

*2.4. Net Ecosystem Production Estimates*

Fluxes of each carbon phase were added up to determine total carbon export, which was then compared to an estimate of net ecosystem production (NEP) of the entire watershed.

---

[1] ftp://aftp.cmdl.noaa.gov/products/trends/co2/co2_mm_mlo.txt (10.06.2017)
[2] http://www.gkd.bayern.de > Abfluss (10.05.2017)
[3] http://www.umweltatlas.bayern.de > Grundlagendaten Fließgewässer (10.05.2017)

MODIS (moderate-resolution imaging spectroradiometer) NPP data of 0.1° resolution, averaged by month and presented in units of g C m$^{-2}$ day$^{-1}$, were used to determine the latter (Zhao et al., 2005). These NPP values were multiplied by the area of the Schwabach Catchment in order to obtain an estimate of the total amount of carbon sequestered annually by the entire watershed. While NPP is an important indicator of the terrestrial carbon uptake of an ecosystem, it does not account for carbon losses via soil respiration. As soil-respired carbon losses can be significant, an estimate of NEP was assumed to be a more relevant indicator of total carbon sequestration, according to the following equation (Liss, 1973)

$$NEP = NPP - R_h \qquad (4)$$

where NEP was calculated by subtracting heterotrophic soil respiration ($R_h$) from NPP. $R_h$ was calculated according to the method of Raich and Potter (1995), who developed fitted equations relating total soil respiration ($S_R$) and air temperature using a global database of soil $CO_2$ effluxes. From this derived total soil-respiration value, $R_h$ was calculated based on the assumption by Hanson et al. (2000) that heterotrophic soil respiration can make up about 54% of $S_R$ in un-forested environments.

## 3. Results

*3.1. DIC, DOC, and POC Characteristics*

Throughout the entire sampling period, DIC exhibited the highest concentrations (30.5 to 56.9 mg C L$^{-1}$) and the most $^{13}$C-enriched δ$^{13}$C values (-10.4 to -14.6‰) (Fig. 2A and 2B). The latter exhibited values intermediate between those expected solely from marine carbonates, and a 50/50 mixture of soil-respired carbon and marine carbonates (Fig. 2A). DIC concentrations were only weakly and positively correlated with δ$^{13}$C$_{DIC}$ ($r^2 = 0.21$, $p < 0.05$) (Fig. 2E), while both DIC concentrations and $\delta^{13}C_{DIC}$ showed a slight negative relationship with discharge ($r^2 = 0.36$, $p < 0.05$ and $r^2 = 0.19$, $p < 0.05$, respectively) (Fig. 2C and 2D).

DOC showed the next highest concentration values (2.5 to 7.4 mg C L$^{-1}$), paired with more $^{13}$C-depleted $\delta^{13}C_{DOC}$ (-23.3 to -28.3‰) that were largely indistinguishable from terrestrial C3 plant values (Fig. 2A,B). Although $\delta^{13}C_{DOC}$ and DOC concentrations were not correlated, the latter showed a noticeable positive exponential correlation with discharge ($r^2 = 0.56$, $p < 0.05$) (Fig. 2C).

The lowest concentrations (0.4 to 3.4 mg C L$^{-1}$) were observed in the POC dataset, which also exhibited the most $^{13}$C-depleted $\delta^{13}$C values (-26.9 to -30‰) (Fig. 2A and 2B). As with DOC, $\delta^{13}C_{POC}$ signatures were nearly identical to those expected from terrestrial C3 plant material (Fig. 2A). POC concentrations and its corresponding $\delta^{13}C_{POC}$ values showed a positive correlation ($r^2 = 0.37$, $p < 0.001$) (Fig. 2E). However, correlations between the POC data, and either the DOC or DIC datasets were not evident. Moreover, POC concentrations were not correlated with river flow, although $\delta^{13}C_{POC}$ exhibited a weakly positive correlation ($r^2 = 0.14$, $p < 0.05$) (Fig. 2C).

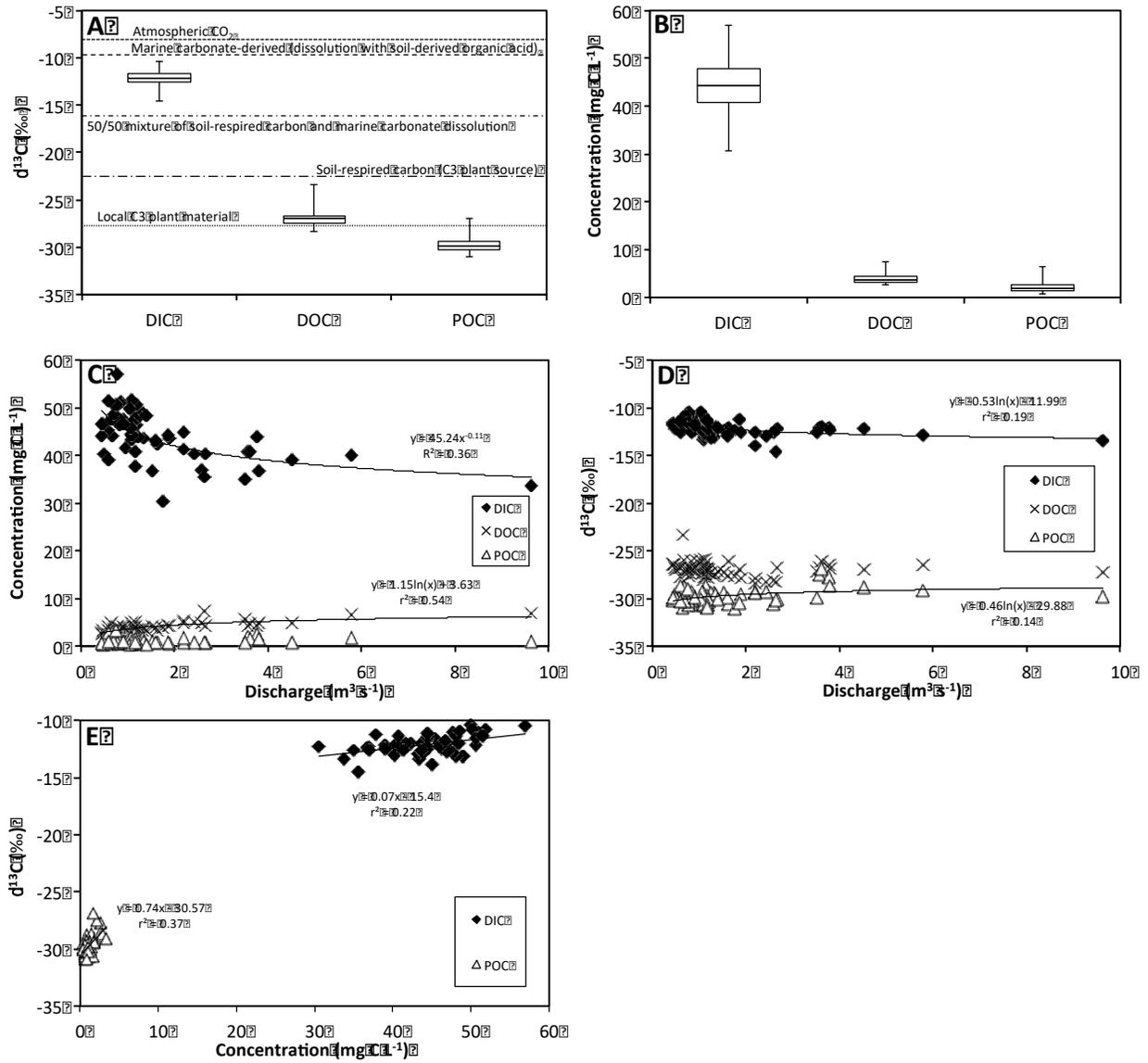

Figure 2. Summary of DIC, DOC, and POC data, including a comparison of (A) $\delta^{13}C$ values, (B) concentration values (whiskers representing maximum and minimum values), (C) cross plots of concentration, (D) $\delta^{13}C$ versus river discharge, and (E) $\delta^{13}C$ versus concentration for DIC and POC.

*3.2. Temporal Variations in Carbon Dynamics and Discharge Relationships*

Figure 3 presents changes as percentage deviations from the yearly averaged value, together with runoff data. It shows that DIC concentrations changes did not follow seasonality patterns and had only comparatively minor variations during the year. The $\delta^{13}C_{DIC}$ data, however, appeared to show more $^{13}C$-depleted values during colder periods (i.e. fall and winter) with a minimum of -20.1‰ during November. Warmer seasons showed more $^{13}C$-enriched $\delta^{13}C_{DIC}$ values with a maximum of -14.2‰ in March (Table 1). DOC concentrations also hardly varied seasonally, however, $\delta^{13}C_{DOC}$ trends were similar to those of $\delta^{13}C_{DIC}$ and showed more $^{13}C$-depleted values in December (minimum of -23.3‰) and $^{13}C$-enriched values in March (maximum of -28.3‰) (Table 1). Neither POC concentrations nor its $\delta^{13}C_{POC}$ values showed any discernible seasonal trends, with comparatively weaker variations when compared to DIC or DOC (Fig. 3).

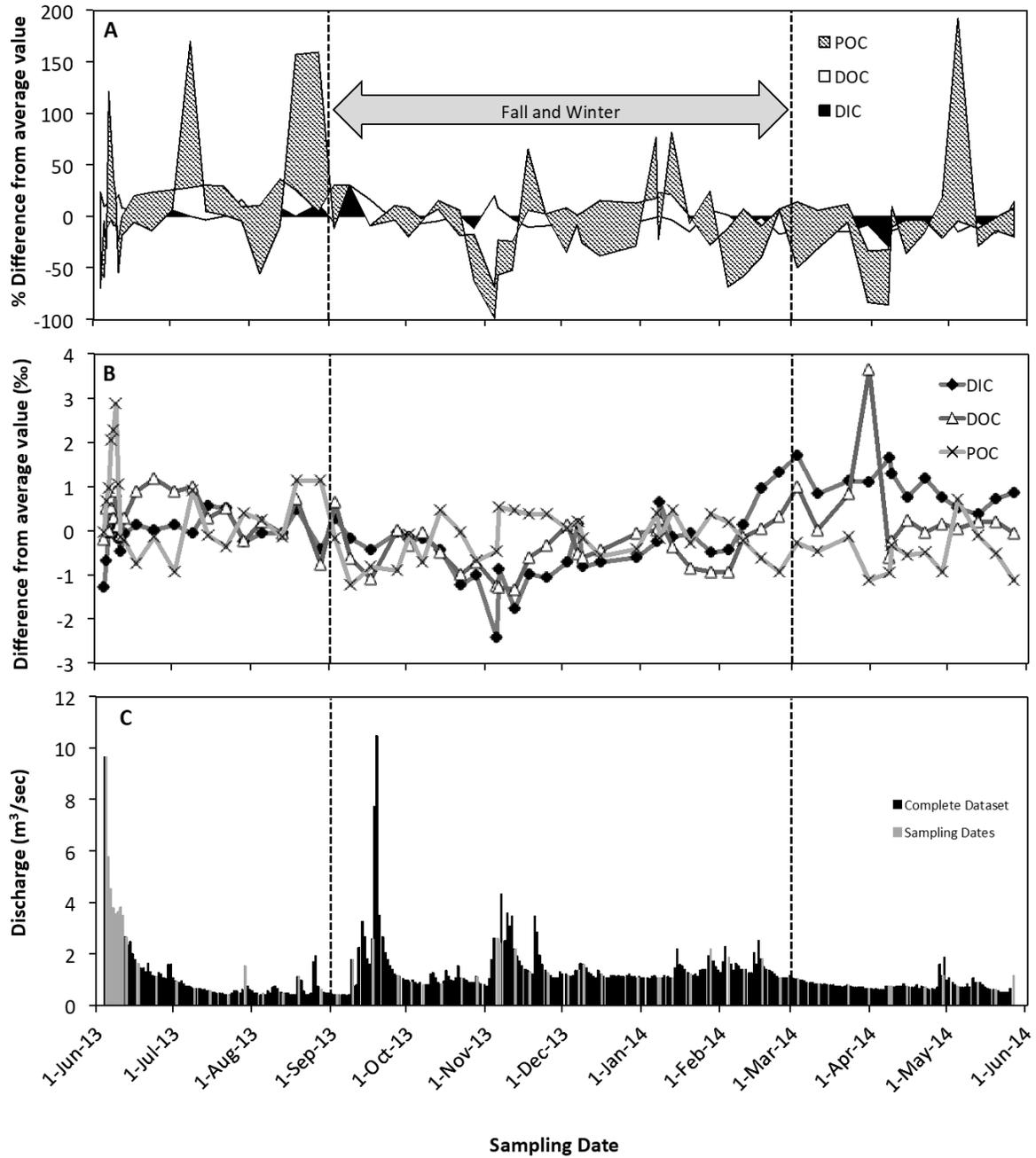

Figure 3. (A) Temporal changes in DIC, DOC, and POC concentration, (B) $\delta^{13}C$ data (shown as ‰ deviations from averaged values), and (C) river discharge during the period of study.

Table 1. Summary of all DIC, DOC, and POC datasets

| Sampling Date | [DIC] (mg C L$^{-1}$) | [DOC] (mg C L$^{-1}$) | [POC] (mg C L$^{-1}$) | $\delta^{13}C_{DIC}$ (‰) | $\delta^{13}C_{DOC}$ (‰) | $\delta^{13}C_{POC}$ (‰) | Discharge (m$^3$ s$^{-1}$) | pCO2 (ppmv) |
|---|---|---|---|---|---|---|---|---|
| 13-06-04 | 33.8 | 7.0 | 1.0 | -13.4 | -27.2 | -29.8 | 9.6 | 1874 |
| 13-06-05 | 40.0 | 6.6 | 1.7 | -12.8 | -26.5 | -29.1 | 5.8 | 1962 |
| 13-06-06 | 39.0 | 4.9 | 0.9 | -12.2 | -27.0 | -28.8 | 4.5 | 2143 |
| 13-06-07 | 43.8 | 4.3 | 2.6 | -12.1 | -26.5 | -27.7 | 3.8 | 3123 |
| 13-06-08 | 40.9 | 4.3 | 2.0 | -12.0 | -26.7 | -27.5 | 3.6 | 2599 |
| 13-06-09 | 40.9 | 4.6 | 1.6 | -11.9 | -26.1 | -26.9 | 3.6 | 2768 |
| 13-06-10 | 36.8 | 4.9 | 1.4 | -12.3 | -26.9 | -28.7 | 3.8 | 2502 |
| 13-06-11 | 35.1 | 5.6 | 0.8 | -12.6 | -27.1 | -29.9 | 3.5 | 2586 |
| 13-06-12 | 40.3 | 4.2 | 0.9 | -12.2 | -26.7 | -30.0 | 2.7 | 4015 |
| 13-06-17 | 42.3 | 3.3 | 0.8 | -12.0 | -26.1 | -30.5 | 1.6 | 2875 |
| 13-06-24 | 43.8 | 3.0 | 0.7 | -12.1 | -25.8 | -29.9 | 1.1 | 2090 |
| 13-07-02 | 41.6 | 3.1 | 0.9 | -12.0 | -26.1 | -30.7 | 1.0 | 2541 |
| 13-07-09 | 44.0 | 2.9 | 2.7 | -12.2 | -26.0 | -28.8 | 0.7 | 4137 |
| 13-07-15 | 45.4 | 2.6 | 0.9 | -11.6 | -26.7 | -29.9 | 0.6 | 4257 |
| 13-07-22 | 44.2 | 2.8 | 0.8 | -11.6 | -26.5 | -30.1 | 0.5 | 3878 |
| 13-07-29 | 36.8 | 4.2 | 1.0 | -12.4 | -27.2 | -29.4 | 1.5 | 3214 |
| 13-08-05 | 46.7 | 3.3 | 0.4 | -12.2 | -26.8 | -29.5 | 0.5 | 3423 |
| 13-08-13 | 40.4 | 2.8 | 0.6 | -12.2 | -27.1 | -29.9 | 0.5 | 4761 |
| 13-08-19 | 44.3 | 2.9 | 2.6 | -11.7 | -26.3 | -28.6 | 1.2 | 4543 |
| 13-08-28 | 39.0 | 4.2 | 2.9 | -12.5 | -27.8 | -28.6 | 0.6 | 2910 |
| 13-09-03 | 46.5 | 2.5 | 0.7 | -11.9 | -26.4 | -29.9 | 0.4 | 2484 |
| 13-09-09 | 30.5 | 4.0 | N/A | -12.3 | -27.6 | -31.0 | 1.8 | 2205 |
| 13-09-17 | 37.1 | 4.9 | N/A | -12.6 | -28.1 | -30.6 | 2.6 | 2239 |
| 13-09-27 | 46.4 | 3.3 | 1.0 | -12.2 | -27.0 | -30.7 | 1.2 | 2785 |
| 13-10-02 | 47.7 | 3.3 | 0.8 | -12.3 | -27.3 | -29.8 | 0.9 | 2216 |
| 13-10-07 | 47.2 | 3.7 | 1.1 | -12.3 | -27.0 | -30.5 | 0.8 | 3343 |
| 13-10-14 | 46.3 | 3.5 | 1.2 | -12.6 | -27.5 | -29.3 | 0.8 | 3195 |
| 13-10-22 | 43.4 | 4.7 | 1.4 | -13.4 | -28.0 | -29.8 | 1.1 | 3834 |
| 13-10-28 | 49.1 | 4.2 | 0.6 | -13.1 | -27.7 | -30.5 | 1.1 | 3959 |
| 13-11-05 | 35.5 | 7.4 | 0.8 | -14.6 | -28.2 | -30.2 | 2.6 | 2306 |
| 13-11-06 | 40.3 | 5.2 | 0.7 | -13.0 | -28.3 | -29.2 | 2.4 | 1489 |
| 13-11-12 | 45.0 | 4.8 | 0.8 | -13.9 | -28.3 | -29.3 | 2.2 | 2471 |
| 13-11-18 | 48.9 | 3.3 | 1.8 | -13.1 | -27.6 | -29.4 | 1.3 | 4119 |
| 13-11-25 | 48.1 | 3.5 | 1.1 | -13.2 | -27.3 | -29.4 | 1.3 | 2379 |
| 13-12-03 | 47.7 | 3.3 | 0.6 | -12.9 | -26.9 | -29.7 | 1.2 | 2295 |
| 13-12-07 | 43.7 | 3.9 | 1.0 | -12.0 | -27.5 | -29.5 | 1.3 | 1906 |
| 13-12-09 | 43.2 | 3.9 | 0.8 | -13.0 | -27.6 | -29.9 | 1.6 | 1897 |
| 13-12-16 | 43.9 | 3.3 | 0.5 | -12.8 | -27.4 | -30.4 | 1.2 | 2384 |
| 13-12-30 | 46.9 | 3.2 | 0.7 | -12.7 | -27.1 | -30.2 | 1.1 | 2459 |
| 14-01-07 | 44.6 | 3.2 | 1.8 | -12.4 | -27.0 | -29.4 | 1.1 | 2635 |
| 14-01-08 | 44.4 | 3.0 | 0.6 | -11.5 | -26.7 | -29.9 | 1.1 | 1985 |
| 14-01-13 | 45.9 | 2.9 | 1.8 | -12.3 | -27.4 | -29.3 | 1.1 | 2807 |
| 14-01-20 | 50.7 | 3.3 | 1.0 | -12.2 | -27.9 | -30.1 | 1.2 | 3435 |
| 14-01-28 | 41.4 | 5.3 | 1.7 | -12.6 | -27.9 | -29.4 | 2.2 | 2769 |
| 14-02-04 | 43.7 | 4.4 | 0.5 | -12.6 | -27.9 | -29.6 | 1.9 | 3074 |
| 14-02-10 | 48.4 | 3.3 | 0.4 | -12.0 | -27.1 | -30.0 | 1.4 | 2656 |
| 14-02-17 | 44.4 | 4.2 | 0.8 | -11.2 | -26.9 | -30.4 | 1.9 | 2610 |
| 14-02-24 | 51.8 | 3.0 | 1.1 | -10.8 | -26.7 | -30.7 | 1.1 | 2743 |
| 14-03-03 | 49.9 | 2.9 | 0.4 | -10.4 | -26.0 | -30.1 | 1.0 | 2208 |
| 14-03-11 | 51.1 | 3.1 | 0.7 | -11.3 | -27.0 | -30.2 | 0.9 | 2849 |
| 14-03-23 | 50.9 | 2.9 | 0.9 | -11.0 | -26.2 | -29.9 | 0.8 | 2991 |
| 14-03-31 | 47.7 | 4.9 | 0.6 | -11.0 | -23.3 | -30.9 | 0.7 | 3240 |
| 14-04-08 | 56.9 | 4.0 | 0.5 | -10.5 | -27.6 | -30.7 | 0.8 | 3239 |
| 14-04-09 | 50.1 | 3.8 | 1.4 | -10.8 | -27.2 | -30.1 | 0.8 | 2762 |
| 14-04-15 | 48.2 | 3.7 | 0.8 | -11.4 | -26.8 | -30.3 | 0.8 | 4368 |
| 14-04-22 | 48.6 | 3.7 | 1.0 | -10.9 | -27.0 | -30.3 | 0.7 | 2672 |
| 14-04-29 | 40.7 | 5.0 | 1.6 | -11.4 | -26.8 | -30.7 | 1.2 | 3645 |
| 14-05-05 | 50.6 | 3.6 | 3.4 | -11.6 | -27.0 | -29.1 | 0.7 | 3097 |
| 14-05-13 | 46.7 | 4.2 | 0.9 | -11.8 | -26.8 | -29.9 | 0.9 | 3820 |
| 14-05-20 | 51.5 | 3.3 | 1.0 | -11.4 | -26.8 | -30.3 | 0.6 | 4111 |
| 14-05-27 | 37.8 | 4.2 | 0.8 | -11.3 | -27.1 | -30.9 | 1.2 | 4266 |

*3.3. Riverine pCO$_2$*

Calculations of $pCO_2$ showed that river waters were over-pressurized in $CO_2$ throughout the entire study period, suggesting that the Schwabach River acts as a net $CO_2$ source to the atmosphere. Relative to ambient atmospheric values, *ep*CO$_2$ (defined as the ratio of riverine $pCO_2$ to atmospheric values; Hope et al. (1994)) values ranged from 3.8 to 12. We found no observable seasonal trend in *ep*CO$_2$ over the sampling period, with variations in *ep*CO$_2$ being apparently unrelated to the time of year (Fig. 4). In addition, epCO$_2$ was not significantly correlated to river discharge, or the DIC, DOC, or POC datasets. These $CO_2$ concentrations corresponded to annually-averaged outgassing rates ranging from 6.84 at minimum *k* values to 13.16 g C m$^{-2}$ d$^{-1}$ at maximum *k* estimates, with a median value of 8.65 g C m$^{-2}$ d$^{-1}$.

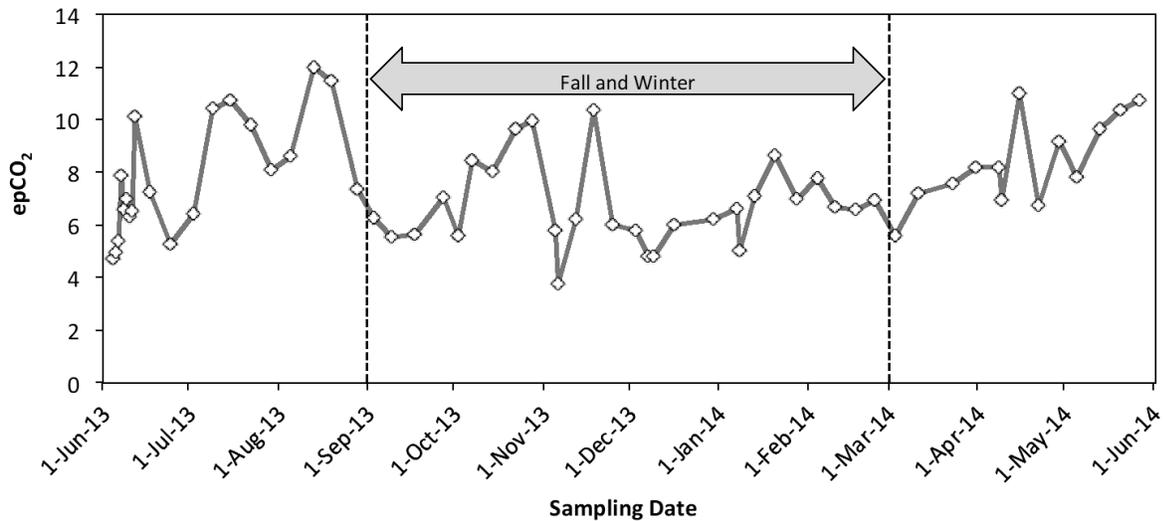

Figure 4: Calculated *ep*CO$_2$ (ratio of $pCO_2$ to ambient atmospheric $CO_2$ level) of the Schwabach River throughout the sampling period

*3.4. Annual Total Carbon Fluxes*

Using all dates of sampling, and incorporating DIC, DOC, POC, and watershed $CO_2$ fluxes, the annual carbon export of the Schwabach Catchment during the study period was calculated to be 2.61 kt C yr$^{-1}$. Of this total export, DIC comprised 70% (1.82 kt C yr$^{-1}$), DOC 7% (0.18 kt C yr$^{-1}$), POC 2% (0.05 kt C yr$^{-1}$), and $CO_2$ efflux 21% (0.56 kt C yr$^{-1}$) (Fig. 5). As all samples were taken from the river mouth, and upstream and tributary data was unavailable for this study, the $CO_2$ fluxes may be underestimated. However, unpublished data retrieved from various sites along the Schwabach River, from its mouth to its source, during an overlapping time period, show a similar $CO_2$ outgassing rate (0.62 kt C yr$^{-1}$) and total carbon flux percentage (28%). With this, we demonstrated that the estimates from the present study are generally indicative of the magnitude of $CO_2$ outgassing from the entire surface of the main stem river.

To determine how total carbon flux estimates may be affected by sampling frequencies, measurement intervals of about 15 days (i.e., bi-monthly) and 30 days (i.e., monthly) were selected, depending on data availability. The bi-monthly scheme was converted into monthly averages, and these were then incorporated into an annually-averaged flux value. Moving averages for both measurement frequencies were calculated in order to determine all possible flux values within the time frame of a year.

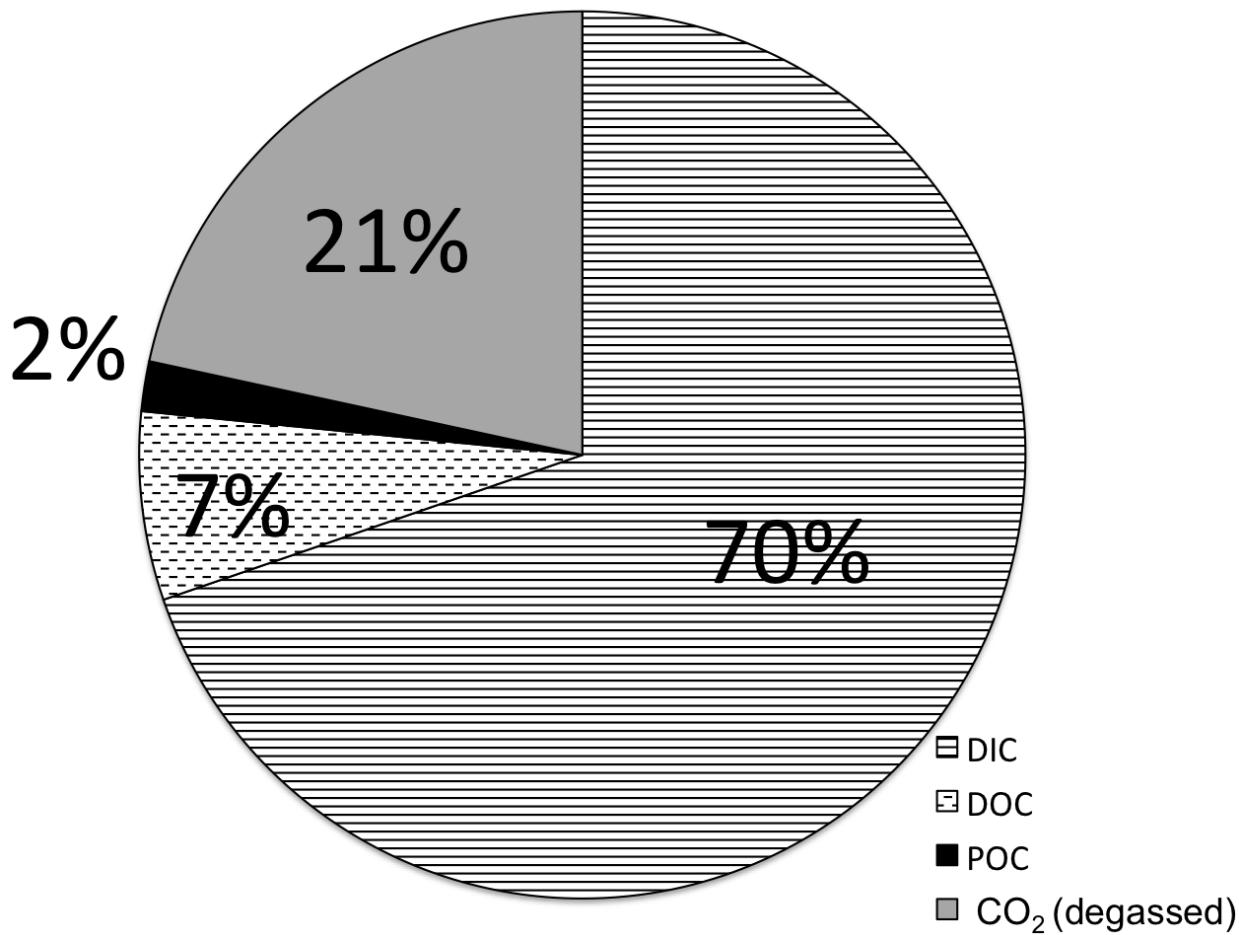

Figure 5. Total annually-averaged flux percentages for each carbon phase.

The data set had a sufficiently high resolution to afford calculations of percent differences between the total carbon flux based on all data, and those calculated by bi-monthly and monthly intervals. The results of this comparison showed that deviations from the complete carbon flux estimates were minor for the bi-monthly scheme, with total carbon flux estimates ranging from a minimum of 2.57 to a maximum 2.69 kt C yr$^{-1}$. This corresponds to percent differences of -1.7 to +3.1% between the two approaches. Deviations from the total carbon flux estimate as based on the complete dataset were much larger with the monthly interval, with fluxes ranging from 2.55 to 3.10 kt C yr$^{-1}$. These correspond to percent differences of -2.3 to +18.8%. (Fig. 6). Moreover, the maximum percent differences between the total carbon flux estimate using all data as compared to that based on monthly intervals was positively correlated to monthly-averaged flow rate ($r^2 = 0.77$, $p < 0.01$), thus indicating that the largest deviations occurred when river flow rates were high. During periods of lower flow rates, percent differences were minor with the selection of either monthly or bi-monthly sampling schemes.

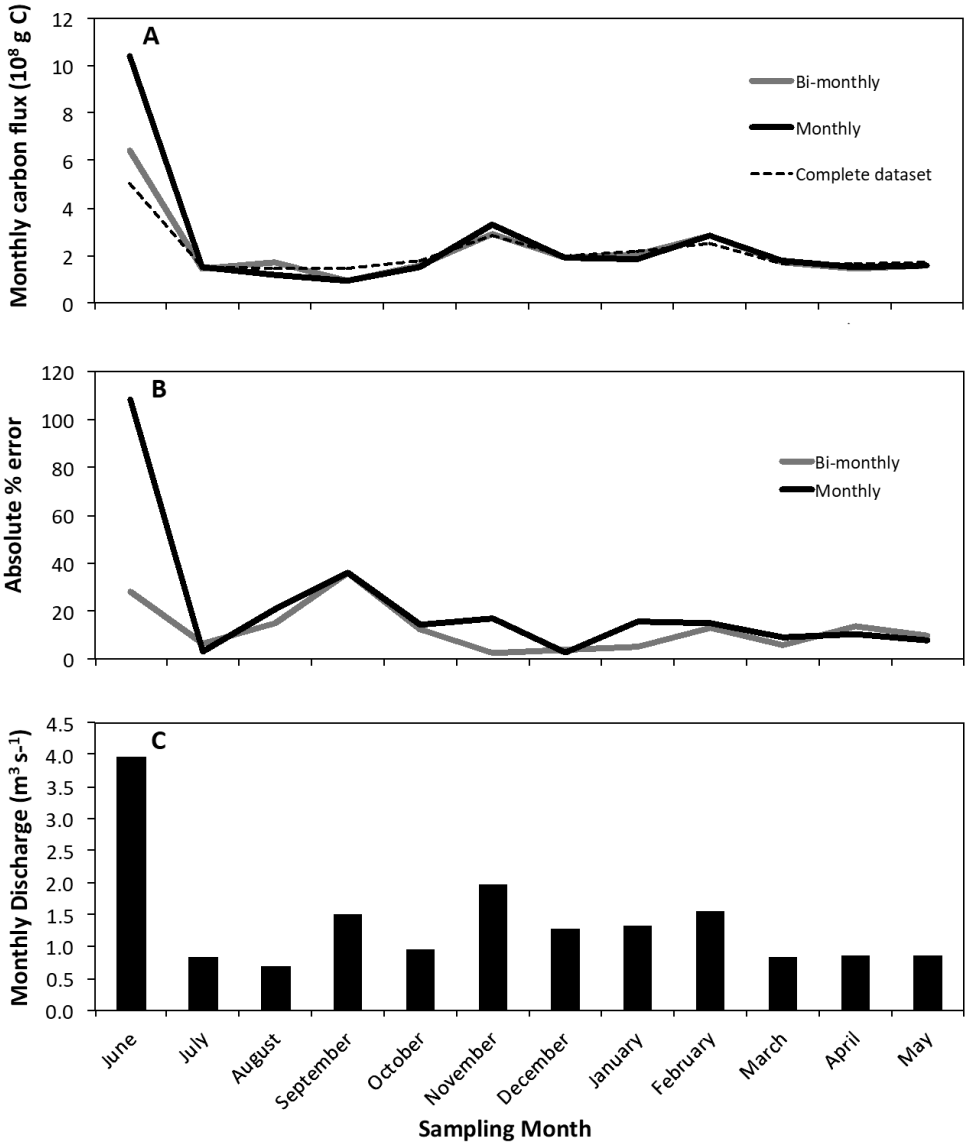

Figure 6. (A) Comparison of total carbon fluxes using the complete dataset, bi-monthly, and monthly measurements (with selection of estimates showing the strongest deviation from those using the complete dataset), (B) percentage deviation of these from the total carbon flux estimate utilizing all data, and (C) discharge data averaged by month

Using the MODIS dataset, an averaged annual NPP of 159.5 kt C yr$^{-1}$ was estimated for the entire watershed. From this value, an $R_h$ of 67.7 kt C yr$^{-1}$ was subtracted to obtain a NEP estimate of 91.8 kt C yr$^{-1}$. Based on this calculation, the total fluvial carbon export is 2.8% of the carbon sequestered by the entire Schwabach Basin.

**4. Discussion**

*4.1. Sampling Frequency*

Our results show that sampling frequency can have a noticeable effect on total carbon flux estimates, particularly those based on monthly sampling schedules. We showed that monthly measurements might be insufficient to capture sudden environmental or climatic changes that can strongly affect riverine carbon flux dynamics. In many cases, the frequency of sampling is influenced by budgetary and logistical constraints, with many studies having necessarily to rely on monthly sampling intervals. However, our calculation of flux errors indicates that samples should at least be taken on a bi-monthly basis for an acceptable threshold of accuracy. This consideration is most significant when accounting for extreme conditions of high or low flow, during which limited sampling could significantly skew flux estimates towards greater errors.

*4.2. Carbon Sources and Sinks*

The DIC dataset indicates a dominance of carbonate lithology as a DIC source, given the comparatively $^{13}$C-enriched $\delta^{13}$C values relative to those expected from silicate weathering (assumed to be largely indistinguishable from the source soil respiration-derived carbonic acid, i.e., roughly -23 ‰ in areas dominated by C3 vegetation; (Cerling et al., 1991). This is further supported by the observed trend of increasing $\delta^{13}C_{DIC}$ with concentration, indicating DIC gain by carbonate dissolution in conjunction with an isotopic shift towards the higher marine carbonate $\delta^{13}$C values of the Upper Jurassic limestone lithology of the catchment. As well, the negative relationship between discharge and DIC concentration suggests a dilution effect during periods of more pronounced runoff. It is likely that smaller amounts of additional plant or soil-derived

DIC were delivered via surface runoff. This would then dilute the existing groundwater signal sourced from marine carbonate-derived DIC.

Previous studies have observed similar trends in carbonate-dominated catchments, where the dominance of carbonate mineral dissolution as a DIC contributor was assumed (Atekwana and Krishnamurthy, 1998; Barth et al., 2003; Cai et al., 2015; Probst et al., 1994). However, although DIC concentration increases may be minor during periods of high water flow, a statistically significant negative correlation between discharge and $\delta^{13}C_{DIC}$ suggests that increased inputs of soil-respired carbon and organic material via soil- and overland flow contributions are sufficient to influence bulk $\delta^{13}C_{DIC}$ towards more negative values.

Assuming the most probable scenario of carbonate dissolution via soil-derived carbonic acid, the DIC derived under open system conditions from this process would be a 50:50 mixture of geological carbonate and soil $CO_2$ (Clark and Fritz, 1997; Deines et al., 1974), which have values of +3‰ and -22.5‰, respectively, based on the work of van Geldern et al. (2015). Associated chemical reactions are:

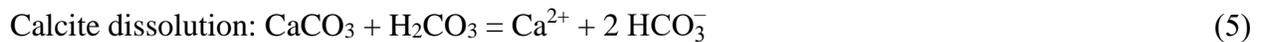

Calcite dissolution: $CaCO_3 + H_2CO_3 = Ca^{2+} + 2\ HCO_3^-$ (5)

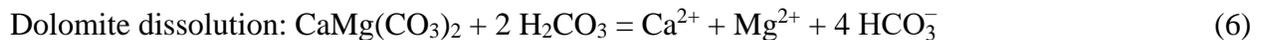

Dolomite dissolution: $CaMg(CO_3)_2 + 2\ H_2CO_3 = Ca^{2+} + Mg^{2+} + 4\ HCO_3^-$ (6)

This corresponds to a $\delta^{13}C_{DIC}$ value of about -9 to -10‰, if the DIC sources consisted solely of dissolved carbonate minerals and soil-derived $CO_2$, and no further alterations of the $^{13}C/^{12}C$ ratio of DIC take place (with $^{13}C$-fractionations from carbonate dissolution assumed negligible in the observed range of river water temperatures; see Salomons and Mook (1986)). Given that the measured $\delta^{13}C_{DIC}$ values are more $^{13}C$-depleted, it is likely that biologically-sourced carbon (probably from the turnover of terrestrially-sourced or in-situ aquatic organic matter) also contributes to the riverine DIC pool.

Even though such biological influences are less pronounced during winter months, they may have caused the more $^{13}$C-depleted DIC values that were observed during this time. The corresponding mechanism of this $^{13}$C-depletion in the river water could be the accumulation of terrestrial organic matter that was subsequently discharged to the river with a time delay, and then transformed into DIC through respiration. Within the river, during the onset of spring and with increasing air temperatures, and thus enhanced primary productivity as compared to respiratory processes, the riverine DIC pool may have become more $^{13}$C-enriched as $^{12}$C was preferentially taken up by photosynthetic activity.

In contrast to DIC, we found a positive relationship between DOC concentration and discharge, thus indicating that overland and soil flows are important sources of riverine DOC during precipitation events and resulting higher runoff. However, a negligible correlation between $\delta^{13}C_{DOC}$ and discharge indicates the same DOC source regardless of enhanced inputs, as shown by the consistent $\delta^{13}C_{DOC}$ signatures. This demonstrates the general dominance of C3 plants as contributors to the fluvial DOC pool.

POC concentrations were poorly correlated to discharge, thus highlighting a reduced importance of overland flow as a POC contributor. Interestingly, a weak but still significant positive relationship between $\delta^{13}C_{POC}$ and discharge was observed. While overall POC inputs may remain mostly unaffected by precipitation events in this system, the nature of the POC may shift in response to hydrological changes. According to *Bayerisches Landesamt für Statistik* (2017)[4] up to 18% of the total agricultural land area of the Schwabach Catchment can be covered by corn crops that belong to the C4 group of plant types. The comparatively more $^{13}$C-enriched plant matter from cornfields may be more prevalent during rainfall events as compared to inputs from native vegetation that is assumed to consist of primarily C3 plant types. While this input may be insufficient to strongly influence bulk POC concentrations, it could still act to shift $\delta^{13}C_{POC}$ towards slightly more $^{13}$C-enriched values with increasing C4 input.

*4.3 Carbon Fluxes and NEP*

---

[4] https://opendata.bayern.de/detailansicht/datensatz/landwirtschaftszaehlung--kreise--anbau-auf-dem-ackerland--fruchtarten--jahr (10.06.2017)

Given that the total riverine carbon export is only 2.8% of the watershed NEP, the Schwabach Catchment appears to be a significant carbon sink. This small percentage of NEP as fluvial export is comparable to a study by Brunet et al. (2009), who found that DIC, DOC, and $CO_2$ cumulatively made up about 3% of watershed NEP in the Nyong Basin, in Cameroon. As well, Shibata et al. (2005) found that DIC, DOC, and POC fluvial export by a first-order stream in a forested basin was only 2% of the watershed NEP. These carbon export percentages, and those found by this study, are much lower when compared to globally-averaged values, whereby 50% to more than 70% of NEP is estimated to be exported into inland waters and lost either via transport to the ocean, $CO_2$ evasion from water surfaces, or long-term sediment storage through burial (Cole et al., 2007; Hope et al., 1994).

This discrepancy may be due to the inclusion of watersheds in productive tropical areas, wetlands, and those dominated by more heavily weathered soils. As the Schwabach Catchment is based around a smaller stream situated in a temperate climate where soil erosion is comparatively small, a smaller proportion of carbon export would be expected as compared to global estimates that encompass a greater variety of climatic regimes and ecological regions. Moreover in our study we were only able to subtract soil respiration from NPP for NEP estimations, which does not account for other respiration process of all possible organisms (plant, microbe and animal), plus some abiotic oxidation into $CO_2$ from processes such as fire and photo-oxidation (Bertilsson et al., 1999; Randerson et al., 2002). Moreover, with this resolution the MODIS database may have overestimated NPP rates. Together with underestimated $R_h$ rates, this may yield underestimated percentages of carbon export when compared to NEP.

Subtracting the carbon loss from these processes would reduce our calculated NEP value and thus increase the percentage of carbon exported by the Schwabach River. In addition, the amount of sediment storage in the Schwabach River is likely to be negligible because as a headwater stream it has no dams or reservoirs. Therefore, this absence of sedimentary storage would have further reduced the percentage exported by the Schwabach River. Overall our study shows that certain headwater catchments, such as those in temperate regions situated in karst bedrock, may not follow the global assumptions established by Cole et al. 2007.

## 5. Conclusions

With our study, we were able to show that sampling frequencies can have a strong effect on the accuracy of total carbon flux estimates. As compared to total carbon fluxes estimated using the complete dataset, the selection of a monthly sampling frequency showed deviations of up to 19%, while those associated with bi-monthly sampling schemes had deviations of only up to 3%. This finding underscores the importance of sampling frequency in calculating riverine carbon fluxes, especially during periods of high river flow. Indeed, flux estimation errors in the monthly sampling scheme show a significant, positive correlation with river discharge. While sampling resolution is often limited by a combination of financial and logistical constraints, our results suggests that a bi-monthly sampling schedule is recommended for sufficiently accurate flux estimates, at least during periods of strong runoff.

This high-resolution dataset suggested that total annual carbon export, being predominantly composed of DIC and sourced mostly from the dissolution of carbonate minerals, made up only a small proportion of NEP in the Schwabach Catchment. Therefore, carbon is largely sequestered by basin vegetation and soils within the Schwabach Watershed. This carbon export is a lower proportion of NEP when compared to global estimates, which may be a product of the ecology and climate of the Schwabach Watershed or uncertainties in the estimation of NEP.

In light of these findings, the authors recommend the application of similarly combined comprehensive DIC, DOC, and POC analyses with high sampling frequencies to watersheds in contrasting and diverse environments. These can be incorporated into existing research initiatives, such as the Global Rivers Observatory, Arctic Great Rivers Observatory, EuroRun, and the Global Network of Isotopes in Rivers (GNIR), for improved and more complete analyses of riverine carbon cycling worldwide.

The present research can therefore be treated as an exploratory study with much potential for growth. Such work can also be expanded by a more comprehensive watershed sampling scheme inclusive of all tributaries and connected springs, and the pairing of stable isotope and concentration analyses with the use of optical laser-based or infrared sensors capable of continuous monitoring (Bastviken et al., 2015; Lorke et al., 2015), providing additional constraints on the accuracy of watershed carbon flux estimates.

**Acknowledgements**

This work was made possible through funding by the Marie Skłodowska-Curie actions (MSCA) via the European Commission [grant number 706088] and the German Research Foundation [grant number BA 2207/6-1]. The authors would like to thank Christian Hanke, Irene Wein, Silke Meyer, and Melanie Hertel for their assistance with geochemical analyses and sample collection.